\documentclass{amsart}
\usepackage{exscale,times}
\usepackage{graphicx}

\DeclareSymbolFont{boperators}{OT1}{cmr}{bx}{n}
\DeclareMathAccent{\karika}{\mathalpha}{boperators}{"17}

\newcommand{\be}{\begin{equation}}
\newcommand{\en}{\end{equation}}
\newcommand{\ee}{\end{equation}}
\newcommand{\bee}[1]{\begin{equation}\label{#1}}
\newcommand{\bey}{\begin{eqnarray}}
\newcommand{\byy}[1]{\begin{eqnarray}\label{#1}}
\newcommand{\eey}{\end{eqnarray}}
\newcommand{\beo}{\begin{eqnarray*}}
\newcommand{\eeo}{\end{eqnarray*}}
\newcommand{\R}[1]{(\ref{#1})}

\newcommand{\beg}{\begin{eqnarray}}
\newcommand{\ene}{\end{eqnarray}}
\newcommand{\mvec}[1]{\mbox{\boldmath{$#1$}}}

\newcommand{\n}{\mvec{n}}

\newcommand{\de}{\mvec{d}}
\newcommand{\eps}{\mvec{\epsilon}}
\newcommand{\D}{\mvec{D}}

\newcommand{\ep}{\epsilon}

\newcommand{\irr}[1]{\hspace{0.2em}
  \stackrel{
   \setbox0=\hbox{\hspace{0.06em}$\displaystyle #1$\hspace{0.06em}}
   \setbox1=\hbox{\vrule width\wd0 height0.08ex depth0pt}
   \vrule width0.08ex height0.08ex depth0.475ex \box1
   \vrule width0.08ex height0.08ex depth0.475ex }{#1}
  \hspace{0.2em}}

\begin{document}
\title{Scalar, vectorial and tensorial damage parameters from the mesoscopic background }
\author{C. Papenfuss$^1$, P. V\'{a}n$^2$}
\address{$^1$ Technische Universt\"at Berlin, Strasse des 17. Juni 135, 10623 Berlin \\ and
\\ Technische Fachhochschule Berlin, Luxemburger Str. 10, 13353 Berlin, Germany}
 \email{c.papenfuss@gmx.de}
 \address{$^2$ KFKI Research Institute for Particle and Nuclear Physics, H-1525, Budapest, P.O. Box 49,
 Hungary}
 \email{vpet@rmki.kfki.hu}

\begin{abstract}In the mesoscopic theory a distribution of different
crack sizes and crack orientations is introduced. A scalar damage
parameter, a second order damage tensor and a vectorial damage
parameter are defined in terms of this distribution function. As an
example of a constitutive quantity the free energy density is given
as a function of the damage tensor. This equation is reduced in the
uniaxial case to a function of the damage vector and in case of a
special geometry to a function of the scalar damage parameter.
\end{abstract}


\maketitle

\section{Introduction}

\subsection{Phenomenological definitions of damage parameters}

Numerous damage models have incorporated scalar, vectorial or
tensorial damage variables that can be characterized at the
macro-scale, for example, by the change in compliance. A scalar
damage parameter has been introduced
\cite{Kac58a,LEM96,KAC80,MUR83} to account for the decrease in the
stiffness of the material with progressing damage. Two scalar
damage parameters have been proposed \cite{JU89} to account
independently for the change in hydrostatic energy and the
remaining part of the elastic energy with increasing damage. A
different reason for introducing two scalar damage parameters was
in \cite{FreNed96} to account for healing of cracks under
compression.  In composites and fiber reinforced materials it is
reasonable to introduce independent scalar order parameters for
the prescribed directions, given by the fiber orientation.

A second order damage tensor has been defined \cite{MUR80},
accounting for the reduction of the effective surface area, which
transmits forces. The resulting effective stress is expressed in
terms of the damage tensor \cite{MUR80}.  For definitions of a
second order damage tensor see also
\cite{KAC80,LEC80,Cha93a,Kan84a,RizCar01}. For parallel
microcracks a second order damage tensor has been associated with
the dyadic product of crack orientation with itself times a scalar
parameter \cite{Betten_aniso}. This definition coincides with our
definition from the mesoscopic point of view in the special case
of parallel microcracks.

A fourth order damage tensor has been introduced. It can be
understood as mapping the elastic tensor of the virgin material to
the elastic tensor of the damaged material, or as mapping the
respective stress tensors. For a summary of damage parameters of
different orders see also \cite{MaiCha97,CauTe99}.

For a constitutive theory of damaged materials with a
thermodynamic background see \cite{MurKam97a,Lem96b}. A
thermodynamic theory of damage, including the interpretation of
failure as loss of thermodynamic stability, can be found in
\cite{Van01a1}. For a comparison to experimental results see
\cite{VanVas01p}.

An alternative choice of damage variable is one that incorporates
salient aspects of damage morphology in its definition. Such
``micro-mechanically- inspired'' damage models involving scalar,
tensor, or ``Fabric tensor'' representations of damage have been
introduced in the study of heterogeneous materials containing
voids or various crack-like surface discontinuities
\cite{NemHor93b,KraFon81a,Kra96b,KraSil82a,Luo_et_al_03,Riz03}.

Our aim here is to show how damage parameters of different tensor
order can be defined from the mesoscopic background. The different
damage parameters correspond to different levels of macroscopic
approximation of the mesoscopic distribution of crack sizes and
orientations. As it has been shown in \cite{TRECOPE04,PaBoeHer07}
in the case of a scalar damage parameter, the mesoscopic theory
leads not only to the definition of damage parameters, but also to
equations of motion for them. On the example of the free energy
density we will show the general form of a constitutive equation
for the different choices of a damage parameter. In the case of a
rotation symmetry of crack orientations, the different forms of
constitutive equation can be reduced to a form with two scalar
parameters: the average crack size and a scalar orientational
order parameter.

\subsection{Mesoscopic theory of complex materials and application to material damage}

The mesoscopic theory has been developed in order to deal with
complex materials within continuum mechanics \cite{Bam}.
 The idea is to enlarge the domain of the
field quantities by an additional variable, characterizing the
internal degree of freedom connected with the internal structure
of the material.  In a simple model the micro-crack is described
as a flat, rotation symmetric surface, a so called penny shaped
crack. In addition we make here the following  simplifying
assumptions:
\begin{enumerate}
\item The diameter of  the cracks is much smaller than the linear
dimension of the continuum element. Under this assumption the
cracks can be treated as an  internal structure of the continuum
element. The cracks are assumed small enough that there is a whole
distribution of crack sizes and orientations in the volume
element. \item The  cracks are fixed to the material. Therefore
their motion is coupled to the motion of representative volume
elements. \item The cracks cannot rotate independently of the
material, i. e. the rotation velocity is determined by the
antisymmetric part of the time derivative of the deformation
gradient of the surrounding material.  \item The number of cracks
is fixed, there is no production of cracks, but very short cracks
are preexisting in the virgin material. \item The cracks cannot
decrease  area, but can only enlarge,  meaning that cracks cannot
heal.
\end{enumerate}
To summarize our model  the micro-crack is characterized by a unit
vector $\n$ representing the orientation of the surface normal and
by the radius  $l$ of the spherical crack surface. These
parameters will be taken as the additional  variables in the
mesoscopic theory.

Beyond the use of additional variables  the mesoscopic concept
introduces a statistical element, the so-called mesoscopic
distribution function. In our case this is a distribution of crack
lengths and orientations in the continuum element at position
$\mvec{x}$ and time $t$, called here crack distribution function
(CDF). The distribution function is the probability density of
finding a crack of length $l$ and orientation $\n$ in the
continuum element. The elements are material elements, including
the same material and the same cracks for all times.  Macroscopic
quantities are calculated from mesoscopic ones as averages  over
crack sizes and crack orientations.

\subsection{Mesoscopic balance equations}

Field quantities such as  mass density, momentum density, angular
momentum density, and energy density are defined on the mesoscopic
space. For distinguishing these fields from the macroscopic ones
we add the word ``mesoscopic``. In addition to mass density we
introduce the crack number density $N$ as the density of  an
extensive quantity. The mesoscopic crack number density $ N (l,
\n, \mvec{x}, t)$
 is the number density, counting only  cracks of length $l$ and orientation
$\n$.

{\em Balance of crack number} \\
In our model the cracks move together with the material element.
Therefore their flux is the convective flux, having a part in
position space, a part in orientation space, and a part in the
length interval. There is no production and no supply of crack
number. Therefore we have for the crack number density $N$:
\bee{C} \frac{\partial}{\partial t}N(\cdot) +
  \nabla_{x}\cdot\{N(\cdot)\mbox{\boldmath{$v$}}(\mvec{x} ,t)\}
  +\nabla_{n}\cdot\{N(\cdot)\mbox{\boldmath{$u$}}(\cdot )\} + \frac{1}{l^2}
  \frac{\partial}{\partial l}
\left( l^2 \dot{l} N(\cdot) \right)\ =\ 0. \ee We have used
spherical coordinates for the mesoscopic variables crack length $l
\in [ 0, \infty ]$ and crack orientation $\mvec{n} \in S^2$, and
we represent the divergence with respect to the mesoscopic
variables in spherical coordinates. $\nabla_{n}$ denotes the
covariant derivative on the unit sphere. $\mvec{v}$ is the
material velocity. In our model all cracks within the continuum
element move with this velocity. $\mvec{u}(\cdot ) =
\dot{\mvec{n}}$ is the orientation change velocity, which is not
the same for all cracks in the continuum element. It is related to
the angular velocity $\mvec{\omega}(\mvec{x} ,t)$ by the relation
\be \mbox{\boldmath{$u$}}(\cdot ) =   \mvec{\omega}\times \mvec{n}
\en This angular velocity is the same for all cracks in the
element. It is determined by the rotation of the surrounding
material.

\subsection{Definition of the distribution function and equation of motion}

Due to its definition as probability density the distribution
function is the number fraction \bee{4} f(l, \n , \mvec{x} ,t)
=\frac{  N(l, \n, \mvec{x} , t)}{ N(\mbox{\boldmath $x$},t)}\quad
, \ee in volume elements, where the number density $N(\mvec{x},t)$
is non-zero.  Here $N(\mbox{\boldmath $x$},t)$ is the macroscopic
number density of cracks of any length and orientation. As the
distribution function in equation \R{4} is not well defined if
$N(\mbox{\boldmath $x$},t) =0$, we define in addition that in this
case $f(l, \n, \mvec{x} ,t)=0$. As there is no creation of cracks
in our model the distribution function will be zero for all times
in these volume elements. In all other volume elements with a
nonzero crack number it is normalized \bee{3}
\int_0^{\infty}\int_{S^2} f(l,\n ,\mvec{x},t)l^2 d^2n dl =1 \quad
. \ee With respect to crack length it is supposed that the
distribution function has a compact  support, meaning that in  a
sample there cannot exist cracks larger than the sample size.

We obtain from the mesoscopic balance of crack number density a
balance of the CDF $f(l,\n, \mvec{x}, t)$,  by inserting its
definition:

\begin{eqnarray}
  \frac{\partial}{\partial t} f(l,\n, \mvec{x}, t) +
  \nabla_{x}\cdot \left( \mbox{\boldmath{$v$}} (\mvec{x}, t) f(l,\n, \mvec{x}, t)\right)
  +\nonumber \\
  \nabla_{n}\cdot \left( \mbox{\boldmath{$u$}} ( \mvec{x}, t) f(l,\n, \mvec{x}, t)\right)
 + \frac{1}{l^2}\frac{\partial }{\partial l}
\left( l^2 \dot{l} f(l,\n, \mvec{x}, t) \right) =0 .\label{f_dgl}
\end{eqnarray}
The right hand side is equal to zero, as for the co-moving
observer the total number of cracks in a volume element does not
change in time.

A growth law for the single crack $\dot{l}$ is needed in equation
(\ref{f_dgl}). For example the {\bf Rice-Griffith dynamics}, which
is motivated from macroscopic thermodynamic considerations, has
been applied.

In  \cite{VanAta00a} the mesoscopic theory has been specialized to
damaged material with penny shaped cracks. The balance equations
and the differential equation for the crack size distribution
function have been derived.  With Rice-Griffith differential
equation for the size of a single crack, the time evolution of the
whole distribution of cracks under load has been investigated, as
well as the evolution of the average crack size \cite{VanPa02}. In
\cite{PaVan03} two different growth laws for the single crack
under load have been considered. Finally, the dynamics of a second
order damage tensor has been derived in \cite{PaBoeHer07} under
the assumption of a simplified single crack growth law under an
effective stress.

\section{Mesoscopic definitions of damage parameters of different orders}
Definitions of a scalar damage parameter, a vectorial parameter
and a damage tensor are given, based on the mesoscopic
distribution function.  A scalar damage parameter
 is the average crack length. The crack growth introduces an anisotropy into the material. In
order to describe this anisotropy, it is necessary to define a
vectorial or a tensorial damage parameter. Starting from the
mesoscopic distribution function, the more natural way is to
define a damage tensor of second order. The tensor character is
introduced by the second moment of the distribution function with
the crack orientation vector.  A case of special interest is a
distribution with a rotation symmetry, the uniaxial case. In this
case the damage tensor can be expressed in terms of a scalar and a
unit vector, which is the orientation of the rotation symmetry
axis. In this case we can define easily  a damage vector from the
second order damage tensor. This damage vector has the orientation
of the rotation symmetry axis. A representation of the second
order damage tensor in the general case without rotation symmetry
is also given. It is shown how a vectorial damage parameter can be
defined in general without rotation symmetry in terms of
eigenvectors of the damage tensor.

\subsection{Damage parameter of second order }

We define the second order damage tensor as the second
orientational moment of the distribution function:

\beg \mvec{D}(\mvec{x} ,t) = <l\irr{\mvec{n}\mvec{n}}> =
\int_0^{\infty} l \int_{S^2} f(l,\mvec{n}, \mvec{x}
,t)\irr{\mvec{n}\mvec{n}}d^2n l^2 {\rm d}l \nonumber \\ =
\int_0^{\infty} \int_{S^2}l^3\irr{\mvec{n}\mvec{n}} f(l,\mvec{n},
\mvec{x} ,t)d^2n {\rm d}l \ene $\irr{\mvec{n}\mvec{n}}$ denotes
the symmetric traceless part of the dyadic product, and $\mvec{D}$
is a second order symmetric traceless tensor.

This definition of the second order damage parameter accounts for
the crack length distribution as well as for the orientation
distribution. With an applied load the length distribution evolves
in time. This shows up in the evolution of the scalar damage
parameter as well as in the evolution of the tensor damage
parameter. As in our model all cracks rotate together with the
surrounding material, the time evolution of the orientation
distribution is a rigid rotation on the unit sphere.

\subsection{Vectorial damage parameter defined from the second
order tensor}

Due to the symmetry the second order tensor damage parameter
$\mvec{D}$ has a spectral decomposition with orthogonal
eigenvectors (unit vectors) $\mvec{d}$, $\mvec{e}$ and $\mvec{f}$,
and eigenvalues $\delta$, $\epsilon$ and $\phi$:

\be \mvec{D} = \delta \mvec{d}\mvec{d} + \epsilon \mvec{e}\mvec{e}
+ \phi \mvec{f} \mvec{f} \ . \en Because $\mvec{D}$ is traceless
we have \be \delta + \epsilon +\phi =0 \ . \en Therefore, not all
eigenvalues can have the same sign. The following cases concerning
the signs of the eigenvalues are possible
\begin{enumerate}
\item One eigenvalue is positive and two eigenvalues are negative,
for instance \be \delta > 0,\quad  \epsilon < 0, \quad \Phi < 0 \
. \en In this case we chose the eigenvector (here $\mvec{d}$)
corresponding to the single positive eigenvalue as the unit vector
defining the orientation of the vector damage parameter. \item One
eigenvalue is negative and two eigenvalues are positive, for
instance \be \delta > 0,\quad  \epsilon > 0, \quad \Phi < 0 \ .
\en In this case we chose the eigenvector (here $\mvec{f}$)
corresponding to the single negative eigenvalue as the unit vector
defining the orientation of the vector damage parameter.
 \item All eigenvalues are zero: \be \delta = 0 , \quad \epsilon =0 , \quad
\phi =0 \ .\en In this case $\mvec{D}=0$ and we have an isotropic
orientation distribution. In this case no vector damage parameter
can be defined. \item One eigenvalue is zero, and the two others
have opposite sign, for instance: \be \delta =0 , \quad \epsilon >
0, \quad \phi = - \epsilon < 0 \ . \en In this very special  case
we  could define a vector damage parameter, having the orientation
of the wedge-product of the two eigenvectors.
\end{enumerate}

The length of the damage vector can be defined as the absolute
value of the corresponding eigenvalue.

The definition of the damage vector in terms of an eigenvector
naturally leads to the symmetry of an orientation, namely the
damage vector and the reversed one cannot be distinguished.

\subsection{Special case of uniaxial distribution function}

If there exists a rotation symmetry axis of the distribution
function, two eigenvalues coincide, either the two positive ones,
or the two negative ones. In both cases the tensor damage
parameter is of the form: \beg \mvec{D} = \langle l S(l) \rangle
\irr{\mvec{d}\mvec{d}}= \langle l S(l) \rangle \left( \mvec{d}
\mvec{d} - \frac{1}{3} \mvec{1}\right)\nonumber \\ = \langle l
S(l) \rangle \left( \mvec{d} \mvec{d} - \frac{1}{3} \left(
\mvec{d} \mvec{d} + \mvec{e} \mvec{e} + \mvec{f} \mvec{f}
\right)\right) \ene with the unit tensor $\mvec{1}$ and a scalar
parameter $S$, denoted as scalar orientational order parameter.
The unit vector $\mvec{d}$ is the orientation of the rotation
symmetry axis. $S(l)$ is a measure of the degree of parallel order
of the cracks. It is zero if the orientations are distributed
isotropically and has the value 1 in the case that all cracks are
oriented parallel. The orientational order can be different for
different crack sizes, therefore $S$ is a function of crack radius
$l$. The average $\langle \quad \rangle $ here is the average over
all crack lengths: \be \langle l S(l) \rangle = \int_0^{\infty}
f(l, \mvec{x},t) l S(l) l^2 dl \ . \en

 For the eigenvalues this corresponds to \beg
\delta =\langle l S(l) \rangle  \left(1- \frac{1}{3} \right) =
\frac{2}{3}  \langle l S(l) \rangle
\\ \epsilon = - \frac{1}{3} \langle l  S(l) \rangle \\ \phi = - \frac{1}{3}  \langle l S(l) \rangle \ . \ene

For positive values of $S$ we have one positive eigenvalue and two
negative ones. For negative values of $S$ two eigenvalues are
positive and one is negative. In both cases the definition of the
vector damage parameter  given in the section above leads to the
eigenvector $\mvec{d}$ as the orientation of the damage vector. In
case of rotation symmetric orientation distributions this is the
orientation of the rotation symmetry axis. The case of positive
values of $S$ corresponds to a distribution, where the
crack-normals are more or less parallel to the rotation symmetry
axis. For negative values of $S$ crack orientations are
concentrated in a plane perpendicular to the rotation symmetry
axis (see figure 1).

\begin{figure}[tbp]
\begin{center}
\includegraphics[height=3cm]{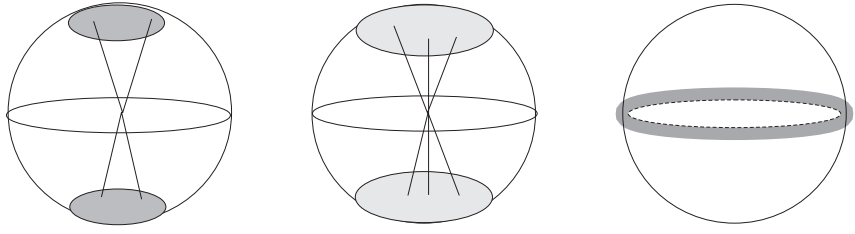}
\end{center}
\caption[CDF]{Different cases of rotation symmetric crack
orientation distributions, corresponding to different values of
the orientational order parameter.  }
\end{figure}

For the damage vector we find in the uniaxial case: \be  \vec{D}
=\frac{2}{3}  \langle l S(l) \rangle \mvec{d} \ .\en

It depends on the degree of orientational order and on the average
crack length.

\subsection{Case of small deviation of the distribution
function from rotation symmetry}

If the deviation of the orientation distribution from rotation
symmetry is small, the two eigenvalues of equal sign differ only
by a small amount $s $, and the damage tensor is of the form: \be
\mvec{D} =  \left( \frac{2}{3} \langle S(l) l \rangle
\mvec{d}\mvec{d} + \left( -\frac{1}{3}\langle ( S(l) -s(l))l
\rangle \right) \mvec{e}\mvec{e} +  \left( -\frac{1}{3} \langle (
S(l) -s(l))l \rangle \right) \mvec{f}\mvec{f} \right)\ .  \en In
this case we can still define the damage vector the same way as in
the uniaxial case: \be \vec{D} =\frac{2}{3} \langle l S(l) \rangle
\mvec{d} \ .\en The scalar order parameter $S$ is a measure of the
degree of order and the biaxiality parameter $s$ is a measure of
the deviation of the orientation distribution from rotation
symmetry.

 \subsection{Scalar damage parameters}
One possible definition of a scalar damage parameter is the
average crack length: \be D = \int_0^{\infty} l f(l,\mvec{x},t)
l^2 dl \ . \en In the rotation symmetric case another scalar
measure of damage is \be D_S = \langle l S(l) \rangle =
\int_0^{\infty} l S(l) f(l,\mvec{x},t) l^2 dl \ . \en It is the
average crack radius, projected onto the plane perpendicular to
the rotation symmetry axis. This is the quantity relevant for
describing the progressive damage under external load in case of
an anisotropic crack distribution.

\section{Examples of constitutive functions for the different
damage descriptions}

In this section we will show on the example of the free energy
density, how the constitutive equation reduces from a
representation in terms of the second order damage tensor to one
in terms of a vector, or a scalar, respectively.

We will assume, that constitutive quantities depend on the
equilibrium variables strain and  temperature and in addition on
the damage parameter. The temperature dependence will not be
denoted explicitly, as it is a scalar quantity. All material
coefficients may depend on temperature.

\subsection{Free energy as a function of strain
and damage in case of a second order damage tensor}

The most general polynomial form of the energy density up to
second order in each variable is given by a representation theorem
 \cite{Smi64a,PipRiv59a}:

 \beg F(\eps , \D, D ) = a_2 D  +
\frac{b_1}{2} D^2 + b_2 tr (\D \cdot  \D )\nonumber \\ + \left(
a_1 + b_4 D +
c_2 D^2 + c_5 tr(\D\cdot \D) \right) tr (\eps) \nonumber \\
\left( b_3 + c_4 D \right) tr (\eps \cdot \D ) + c_8 tr (\D \cdot
\D \cdot \eps) \nonumber \\ + \left( \frac{\lambda}{2} + c_1 D  +
d_1 D^2 + d_3 tr( \D\cdot \D) \right) \left( tr (\eps ) \right)^2
\nonumber
\\ + \left( \mu + c_6 D + d_2 D^2 +d_4 tr (\D \cdot \D ) \right) tr (\eps \cdot \eps ) \nonumber \\
+ \left( c_3 tr ( \eps)+ d_5 tr (\eps \cdot \D ) + d_6 tr(\eps) D
\right) tr(\eps \cdot \D ) \nonumber \\ + \left( c_7 + d_8 D
\right) tr (\eps \cdot \eps \cdot \D) + d_7 tr(\eps) tr(\D\cdot \D
\cdot \eps ) \nonumber \\ + d_9 tr (\D\cdot \D \cdot \eps\cdot
\eps )
\ene

This form is the simplest and natural extension of linear
elasticity considering the damage. The coefficients still can be
arbitrary functions of temperature.

In the case of rotation symmetry we have: \be \D = D_S
\irr{\mvec{d}\mvec{d}} \en and the scalar products can be
calculated as: \beg \D \cdot \D = D_S^2 \irr{\mvec{d}\mvec{d}}
\cdot \irr{\mvec{d}\mvec{d}}\nonumber \\ =  D_S^2
\left(\mvec{d}\mvec{d}- \frac{1}{3} \mvec{1}\right) \cdot
\left(\mvec{d}\mvec{d}- \frac{1}{3} \mvec{1}\right) = D_S^2 \left(
\frac{1}{3} \mvec{d}\mvec{d} + \frac{1}{9} \mvec{1} \right) \ene
and \beg tr (\D\cdot \D )= \frac{2}{3} D_S^2\ene The expression
for the free energy simplifies to: \beg F(\eps , D_S, \mvec{d}, D
) = a_2 D + \frac{b_1}{2} D^2 + \frac{2}{3} b_2 D_S^2 + \nonumber \\
tr (\eps) \left( a_1 + b_4 D + c_2 D^2 - \frac{1}{3} b_3 D_S -
\frac{1}{3} c_4 D D_S + (\frac{2}{3} c_5 + \frac{1}{9} c_8  )^2
D_S^2\right) \nonumber \\ + ( tr\eps)^2 \left( \frac{1}{2} \lambda
+ c_1 D + d_1 D^2 - \frac{1}{3} c_3 D_S - \frac{1}{3}d_6 D D_S
\right. \nonumber \\ \left. + (
 \frac{2}{3} d_3+   \frac{1}{9} d_5  + \frac{1}{9} d_7 ) D_S^2
 \right) + \nonumber \\  tr (\eps \cdot \eps ) \left( \mu + c_6 D
 + d_2 D^2  - \frac{1}{3} c_7 D_S  - \frac{1}{3} c_8 D D_S    + ( \frac{2}{3}
 d_4 +  \frac{1}{9} d_9) D_S^2\right) \nonumber \\
+ \eps : \mvec{d} \mvec{d} \left( b_3 D_S + c_4 D D_S +
\frac{1}{3} c_8 D_S^2 \right) + \left( \eps : \mvec{d}
\mvec{d}\right)^2 d_5 D_S\nonumber \\ + ( tr \eps) \eps : \mvec{d}
\mvec{d} \left(( d_5+ c_3) D_S + d_6 D D_S + \frac{1}{3} d_7 tr
\eps D_S^2\right)\nonumber \\ + \mvec{d}\cdot \eps \cdot \eps
\cdot \mvec{d} \left( c_7 D_S + d_8 D D_S + \frac{1}{3} d_9 D_S^2
\right) \ene

The free energy density is expressed here in terms of the  vector
$\mvec{d}$ (the rotation symmetry axis) and the scalar damage
parameter $D_S$.

\subsection{Special case of uniaxial strain in the z-direction and
symmetry axis of the distribution in the same direction}

This situation occurs (approximately) in a uniaxial tension
experiment (see figure 2).

\begin{figure}[tbp]
\begin{center}
\includegraphics[height=4cm]{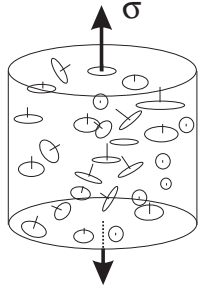}
\end{center}
\label{exp} \caption{Schematic view of an experiment with uniaxial
loading. } \end{figure}

The assumption that in all volume elements the CDF is rotation
symmetric with the $z$-direction as symmetry axis is an
approximation, valid for small deformations. In case of large
deformations, the rotation of the volume element cannot be
neglected, and the cracks rotate with the material element. This
leads to a rotation of the symmetry axis of the orientation
distribution function in the volume element, which depends on the
position of the volume element. The (local) symmetry axis of the
CDF does not coincide with the direction of the applied strain
anymore.

In the geometry with the global rotation symmetry around the
z-axis the only interesting components of tensors are the
z-z-components and traces. For the strain and the damage vector we
have:

\beg \eps = \epsilon_{zz} \mvec{e}_z \mvec{e}_z =: \epsilon
\mvec{e}_z \mvec{e}_z  \\ \de = \mvec{e}_z  \ene and the free
energy density reduces to a function of three scalar quantities
$\epsilon, D, D_S$. $D$ is the average crack length (in any
direction) and $D_S$ is a measure of the anisotropy of the average
crack length distribution in the z-direction and orthogonal to it.

\beg F(\ep,D_S, D) = a_2 D + \frac{b_1 }{2} D^2 + \frac{2}{3} b_2
D_S^2 \nonumber \\ + \ep \left( a_1 + b_4 D + c_2 D^2 + \left(
\frac{2}{3} b_3 + \frac{2}{3}c_4 D\right)D_S + \left( \frac{2}{3}
c_5 + \frac{4}{9}c_8 \right) D_S^2 \right)  \nonumber
\\ \ep^2  \left( \frac{\lambda }{2} + \mu +\left( c_1 + c_6\right) D
+ \left( d_1 + d_2\right) D^2 \right . \nonumber \\ D_S \left(
\frac{2}{3}c_3 + \frac{2}{3} c_7 + D \left( \frac{2}{3} d_6
 +  \frac{2}{3} d_8\right)\right) \nonumber \\  \left. + D_S^2 \left( \frac{2}{3} d_3+ \frac{2}{2} d_4 + \frac{4}{9} d_5+ \frac{4}{9}
 d_7 + \frac{4}{9} d_9 \right)\right) \ep^2 \ene

\section{Acknowledgements}

 C. Papenfuss thanks  the Deutsche
Forschungsgemeinschaft (DFG), contract PA 410/5-1, for financial
support.  This research was supported by OTKA T048489, by the
EU-I3HP project and the Bolyai scholarship of Peter V\'an.



\begin{thebibliography}{10}

\bibitem{Kac58a}
L.~M. Kachanov.
\newblock On the time to failure under creep conditions.
\newblock {\em Izv. AN SSSR, Otd. Tekhn. Nauk.}, (8):26--31, 1958.

\bibitem{LEM96}
J.~Lemaitre.
\newblock {\em A {C}ourse on {D}amage {M}echanics}.
\newblock Springer-Verlag, Berlin, New York, Heidelberg.

\bibitem{KAC80}
M.~Kachanov.
\newblock Continuum model of medium with microcracks.
\newblock {\em J. of the Engeneering Mechanics division}, 106(EM5):1039--1051,
  1980.

\bibitem{MUR83}
S.~Murakami.
\newblock Notion of {C}ontinuum {D}amage {M}echanics and its {A}pplication to
  {A}nisotropic {C}reep {D}amage {T}heory.
\newblock {\em J. Engng. Mat. Technology}, 105:99, 1983.

\bibitem{JU89}
J.W. Ju.
\newblock On energy-based coupled elastoplastic damage theories: {C}onstitutive
  modelling and computational aspects.
\newblock {\em Int. J. Solid Structures}, 25(7):803--833, 1989.

\bibitem{FreNed96}
M.~Fr\'{e}mond and B.~Nedjar.
\newblock Damage, gradient of damage and principle of virtual power.
\newblock {\em Intl. J. Solids Structures}, 33(8):1083--1103, 1996.

\bibitem{MUR80}
S.~Murakami and N.~Ohno.
\newblock A continuum theory of creep and creep damage.
\newblock In D.~R. Ponter, A. R.~Hayhorst, editor, {\em 3rd {C}reep in
  {S}tructures {S}ymposium, {L}eicester}, pages 422--443, Berlin, New York,
  Heidelberg, 1980. IUTAM, Springer.

\bibitem{LEC80}
F.~A. Lecki and E.~T. Onat.
\newblock Physical {N}onlinearities in {S}tructural {A}nalysis.
\newblock In {\em Tensorial nature of damage measuring internal variables},
  Berlin, New York, Heidelberg, 1980. Springer-Verlag.

\bibitem{Cha93a}
J.-L. Chaboche.
\newblock Development of continuum damage mechanics for elastic solids
  sustaining anisotropic and unilateral damage.
\newblock {\em International Journal of Damage Mechanics}, 2:311--329, 1993.

\bibitem{Kan84a}
K.-I. Kanatani.
\newblock Distribution of directional data and fabric tensors.
\newblock {\em International Journal of Engineering Science}, 22(2):149--164,
  1984.

\bibitem{RizCar01}
E.~Rizzi and I.~Carol.
\newblock A formulation of anisotropic elastic damage using compact tensor
  formalism.
\newblock {\em J. OF ELASTICITY}, 64(2-3):85--109, 2001.

\bibitem{Betten_aniso}
J.~Betten, A.~Sklepus, and A.~Zolochevsky.
\newblock A constitutive theory for creep behavior of initially isotropic
  materials sustaining unilateral damage.
\newblock {\em Mechanics Research Communications}, 30:251–256, 2003.

\bibitem{MaiCha97}
J.~F. Maire and J.~L. Chaboche.
\newblock A new formulation of continuum damage mechanics (cdm) for composite
  materials.
\newblock {\em Aerospace Science and Technology}, 4:247--257, 1997.

\bibitem{CauTe99}
A.~Cauvin and Testa R.B.
\newblock Damage mechanics: basic variables in continuum theories.
\newblock {\em INT. J. SOLIDS AND STRUCTURES}, 36(5):747--761, 1999.

\bibitem{MurKam97a}
S.~Murakami and K.~Kamiya.
\newblock Constitutive and damage evolution equations.
\newblock {\em Int. J. Mech. Sci.}, 39(4):473--486, 1997.

\bibitem{Lem96b}
J.~Lemaitre.
\newblock {\em A Course on Damage Mechanics}.
\newblock Springer Verlag, Berlin-etc., 1996.

\bibitem{Van01a1}
P.~V\'an.
\newblock Internal thermodynamic variables and the failure of microcracked
  materials.
\newblock {\em Journal of Non-Equilibrium Thermodynamics}, 26(2):167--189,
  2001.

\bibitem{VanVas01p}
P.~V\'an and B.~V\'as\'arhelyi.
\newblock Second law of thermodynamics and the failure of rock materials.
\newblock In J.~P.~Tinucci D.~Elsworth and K.~A. Heasley, editors, {\em Rock
  Mechanics in the National Interest V1}, pages 767--773, Lisse-Abingdon-
  Exton(PA)-Tokyo, 2001. Balkema Publishers.
\newblock Proceedings of the 9th North American Rock Mechanics Symposium,
  Washington, USA, 2001.

\bibitem{NemHor93b}
S.~Nemat-Nasser and M.~Hori.
\newblock {\em Micromechanics: Overall Properties of Heterogeneous Materials}.
\newblock North-Holland, Amsterdam, 1993.

\bibitem{KraFon81a}
D.~Krajcinovic and G.~U. Fonseka.
\newblock The continuous damage theory of {brittle} materials, {P}art 1:
  General theory.
\newblock {\em Journal of Applied Mechanics}, 48:809--815, 1981.

\bibitem{Kra96b}
D.~Krajcinovic.
\newblock {\em Damage mechanics}.
\newblock Elsevier, Amsterdam-etc., 1996.
\newblock North-Holland Series in Applied Mathematics and Mechanics.

\bibitem{KraSil82a}
D.~Krajcinovic and M.~A.~G. Silva.
\newblock Statistical aspects of the continuous damage mechanics.
\newblock {\em International Journal Solids Structures}, 18:551--562, 1982.

\bibitem{Luo_et_al_03}
D.M. Luo and K.~Takezono, S.and~Tao.
\newblock The mechanical behavior analysis of cfcc with overall anisotropic
  damage by the micro-macro scale method.
\newblock {\em Int. J. Damage Mechanics}, 12(2):141--162, 2003.

\bibitem{Riz03}
E.~Rizzi and I.~Carol.
\newblock Dual orthotropic damage-effect tensors with complementary structures.
\newblock {\em Int. J. Engn. Sci.}, 41(13-14):1445--1495, 2003.

\bibitem{TRECOPE04}
C.~Papenfuss.
\newblock Damage evolution in micro-cracked materials under load.
\newblock In B.T. Maruszewski, W.~Muschik, and A.~Radowicz, editors, {\em
  Trends in Continuum Physics}. World Scientific, 2004.

\bibitem{PaBoeHer07}
C.~Papenfuss, T.~B\"ohme, H.~Herrmann, W.~Muschik, and
J.~Verh\'{a}s. \newblock Dynamics of the Size and Orientation
Distribution of Microcracks and Evolution of Macroscopic
  Damage Parameters. \newblock {\em J. Non-Equilib. Thermodyn. },
 32(2): 129--143, 2007.

\bibitem{Bam}
C.~Papenfuss.
\newblock Theory of liquid crystals as an example of mesoscopic continuum
  mechanics.
\newblock {\em Computational Materials Science}, 19:45 -- 52, 2000.

\bibitem{VanAta00a}
P.~V\'an, C.~Papenfuss, and W.~Muschik.
\newblock Mesoscopic dynamics of microcracks.
\newblock {\em Phys. Rev. E}, 62(5):6206--6215, 2000.

\bibitem{VanPa02}
P.~V\'an, C.~Papenfuss, and W.~Muschik.
\newblock Griffith cracks in the mesoscopic microcrack theory.
\newblock {\em J. Phys. A}, 37(20):5315--5328, 2004.
\newblock published online: Condensed Matter, abstract, cond-mat/0211207; 2002.

\bibitem{PaVan03}
C.~Papenfuss, P.~V\'an, and W.~Muschik.
\newblock Mesoscopic theory of microcracks.
\newblock {\em Archive of Mechanics}, 55(5-6):481--499, 2003.

\bibitem{Smi64a}
G.~F. Smith.
\newblock On isotropic integrity bases.
\newblock {\em Archive for Rational Mechanics and Analysis}, 17:282--292, 1964.

\bibitem{PipRiv59a}
A.~C. Pipkin and R.~S. Rivlin.
\newblock The formulation of constitutive equations in continuum physics 1.
\newblock {\em Archive for Rational Mechanics and Analysis}, 4:129--144, 1959.

\end{thebibliography}

\end{document}